\documentclass[12pt,preprint]{aastex}

\usepackage{multirow}

\shorttitle{C/2007 N3 (Lulin) observed with {\it AKARI}}
\shortauthors{Ootsubo et al.}

\begin{document}


\title{Detection of parent H$_2$O and CO$_2$ molecules in the 2.5--5 $\mu$m spectrum of 
comet C/2007 N3 (Lulin) observed with {\it AKARI}}


\author{Takafumi Ootsubo\altaffilmark{1,2}, 
		Fumihiko Usui\altaffilmark{1}, 
		Hideyo Kawakita\altaffilmark{3},
		Masateru Ishiguro\altaffilmark{4}, 
		Reiko Furusho\altaffilmark{5}, 
		Sunao Hasegawa\altaffilmark{1}, 
		Munetaka Ueno\altaffilmark{1},
		Jun-ichi Watanabe\altaffilmark{5},
		Tomohiko Sekiguchi\altaffilmark{6},
		Takehiko Wada\altaffilmark{1},
		Youichi Ohyama\altaffilmark{7},
		Shinki Oyabu\altaffilmark{1},
		Hideo Matsuhara\altaffilmark{1}, 
		Takashi Onaka\altaffilmark{8},
		Takao Nakagawa\altaffilmark{1}, and 
		Hiroshi Murakami\altaffilmark{1}
		}


\altaffiltext{1}{Institute of Space and Astronautical Science, 
		Japan Aerospace Exploration Agency,
		3-1-1 Yoshinodai, Chuo-ku, Sagamihara, Kanagawa 252-5210, Japan}
\altaffiltext{2}{Astronomical Institute, Graduate School of Science,
		Tohoku University, Aramaki, Aoba-ku, Sendai, 980-8578, Japan, ootsubo@astr.tohoku.ac.jp}
\altaffiltext{3}{Kyoto Sangyo University, 
		Motoyama, Kamigamo, Kita-Ku, Kyoto 603-8555, Japan}
\altaffiltext{4}{Department of Physics and Astronomy, Seoul National University,
		599 Gwanak-ro, Gwanak-gu, Seoul 151-742, Korea}
\altaffiltext{5}{National Astronomical Observatory of Japan, 
		2-21-1 Osawa, Mitaka, Tokyo 181-8588, Japan}
\altaffiltext{6}{Hokkaido University of Education, Asahikawa Campus,
		Hokumon 9, Asahikawa, Hokkaido 070-8621, Japan}
\altaffiltext{7}{Academia Sinica, Institute of Astronomy and Astrophysics,
		PO Box 23-141, Taipei 106, Taiwan}
\altaffiltext{8}{Department of Astronomy, Graduate School of Science, 
		The University of Tokyo,
		7-3-1 Hongo, Bunkyo-ku, Tokyo 113-0033, Japan}


\begin{abstract}
Comet C/2007 N3 (Lulin) was observed with the Japanese infrared 
satellite {\it AKARI} in the near-infrared 
at a post-perihelion heliocentric distance of 1.7 AU.
Observations were performed with the spectroscopic (2.5--5.0 $\mu$m) and 
imaging (2.4, 3.2, and 4.1 $\mu$m) modes on 2009 March 30 and 31 UT, respectively.
{\it AKARI} images of the comet exhibit a sunward crescent-like shape coma and 
a dust tail extended toward the anti-solar direction.
The 4.1 $\mu$m image (CO/CO$_2$ and dust grains) shows a distribution different 
from the 2.4 and 3.2 $\mu$m images (H$_2$O and dust grains).
The observed spectrum shows distinct bands at 2.66 and 4.26 $\mu$m, 
attributed to H$_2$O and CO$_2$, respectively.
This is the fifth comet in which CO$_2$ has been directly detected in the near-infrared spectrum.
In addition, CO at 4.67 $\mu$m and a broad 3.2--3.6 $\mu$m emission
band from C--H bearing molecules were detected in the {\it AKARI} spectrum.
The relative abundance ratios $\rm{CO_2/H_2O}$ and $\rm{CO/H_2O}$ derived from 
the molecular production rates are 
$\sim 4$\%--5\% and $< 2$\%, respectively. 	
Comet Lulin belongs to the group that has relatively low abundances 
of CO and CO$_2$ among the comets observed ever.
\end{abstract}





\section{Introduction}

	Since comets are remnants from the protoplanetary disk and the least processed
	bodies in the solar system, their composition and heterogeneity
	and their relation to their dynamical history help our understanding of the formation process 
	of planetesimals and planets in the early solar nebula.
	One of the main goals of cometary studies is the determination of the 
	composition of volatile species contained in the nucleus as ice.
	Besides water (H$_2$O), carbon monoxide (CO), carbon dioxide (CO$_2$), 
	and methanol (CH$_3$OH) are the most abundant species in comet nuclei.
	They are the key species that provide constraints on a link
    between cometary volatiles and the interstellar precursor \citep{Irvine04, Ehren04, BM04}. 
    There are generally two major dynamical groups for comets: 
    Jupiter family or Ecliptic comets \citep[][and references therein]{Duncan04}
    and Oort cloud comets \citep[][and references therein]{Dones04}. 
	These two groups are expected to have different properties physically or chemically
	due to the difference in their birth places and evolutions. 

	Infrared (IR) spectroscopy is a powerful tool to study the chemical composition
	of comets and to derive information on the formation and evolution 
	mechanisms and on the materials from which they originate.
	Most parent molecules, which sublimate directly from the nucleus, have 
	strong fundamental bands of vibration in the near-IR (2.5--5 $\mu$m).
	H$_2$O and CO$_2$ have the fundamental $\nu_3$ band at 2.66 and 
	4.26 $\mu$m, respectively. CO has the $v$(1--0) band at 4.67 $\mu$m. 
	Although the near-IR region is partly accessible from the ground and
	advances in modern spectrometers with high spectral resolving powers have 
	made a great progress in the study of H$_2$O hot bands and organic molecules
	\citep[][and references therein]{DiSanti08},
	the most abundant parent molecules (especially CO$_{2}$) are difficult to observe
	directly from the ground because they are also present in the terrestrial atmosphere 
	and strong telluric absorption bands make the atmosphere completely opaque.
	Symmetric molecules, such as CO$_2$, do not have a permanent electric dipole 
	moment and thus cannot be observed even in the radio range from the ground.

	The advent of space missions allows us to study the emission from cometary 
	volatiles in the entire 2.5--5 $\mu$m region. 	
	Parent CO$_2$ from the comet nucleus was detected in the coma of comet 1P/Halley 
	by the Russian {\it Vega} space probe for the first time \citep{Combes88}.
	Since then, it has been directly observed in only three other comets: Hale-Bopp 
	\citep{Crovisier96, Crovisier97b,Crovisier99a} and 103P/Hartley \citep{Colangeli99, 
	Crovisier99b} with {\it Infrared Space Observatory} ({\it ISO}), and 9P/Tempel
	with the {\it Deep Impact} flyby spacecraft \citep{AHearn05, Feaga07}.
	With regard to imaging observations, {\it Spitzer Space Telescope} also
	observes the cometary CO and CO$_2$ in the near-IR with the Infrared Array Camera (IRAC).
	Recent studies report the CO$_2$ production rates of comets 21P/Giacobini--Zinner
	and 73P/Schwassmann--Wachmann \citep{Pitticova08, Reach09}.
	The IRAC 4.5 $\mu$m image is, however, a combination of CO, CO$_2$, and dust thermal emission. 
	It is difficult to derive the CO and CO$_2$ production rates separately. 
	
	{\it AKARI}, the Japanese IR satellite \citep{Murakami07}, provides a near-IR 
	spectroscopic capability from the space for the first time after {\it ISO}. 
	We present here the results of a search for parent molecules in comet 
	C/2007 N3 (Lulin), hereafter called C/Lulin, observed with {\it AKARI} at near-IR wavelengths.
	C/Lulin is one of the Oort cloud comets, which was discovered
	by Lulin Observatory (Taiwan) on 2007 July 11.
	The comet passed the perihelion on 2009 January 10.6 UT with the heliocentric distance 
	of 1.21 AU and approached closest to the Earth on February 24 
	with a geocentric distance of 0.41 AU.

\section{Observations and Data Reduction}


	{\it AKARI} is equipped with a 68.5 cm cooled telescope and two scientific instruments, 
	the Far-Infrared Surveyor \citep[FIS;][]{Kawada07} and the Infrared Camera \citep[IRC;][]{Onaka07}. 
	{\it AKARI} was launched on 2006 February 21 UT, and its liquid helium (LHe) cryogen boiled off on 2007 August 26 UT, 
	550 days after launch. 
	In the post-helium phase (Phase 3), 
	the telescope and scientific instruments are kept around 40 K 
	by a mechanical cooler and only near-IR observations (1.8--5.5 $\mu$m) 
	are carried out.	
	Near-IR observations, both imaging and spectroscopy, of C/Lulin 
	were performed with {\it AKARI}/IRC during this post-helium phase as part of the Director's Time observations.	
	We have only several chances to observe this comet because
	{\it AKARI} has the visibility restriction of its solar elongation angle within $90\pm1$ deg.
	At the time of the observation, the comet was at a heliocentric distance of 1.70 AU and
	a geocentric distance of 1.36--1.37 AU.

\begin{figure}
   \centering
   \includegraphics[scale=0.7]{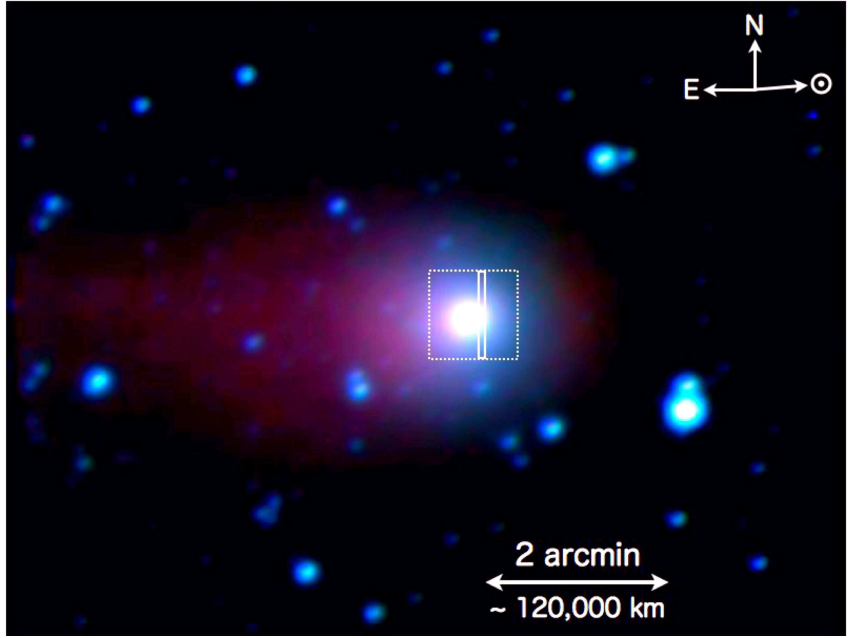}
      \caption{Synthesized false color image of comet C/2007 N3 (Lulin) made from 
      the N2 (blue: 1.9--2.8 $\mu$m), N3 (green: 2.7--3.8 $\mu$m), and N4 (red: 3.6--5.3 $\mu$m) band 
      images obtained with the {\it AKARI}/IRC on 2009 March 31. 
      The schematic aperture position for
      the NG spectroscopy on March 30 is overlaid in the same figure.
      The IRC $1\arcmin \times 1\arcmin$ aperture region (Np) is shown with 
      dotted lines. The area shown with the solid lines (60\arcsec$ \times $4\farcs5 region)
      corresponds to the region where the spectrum of C/Lulin is extracted. 
              }
         \label{Lulin_img}
   \end{figure}

	Spectroscopic observations were carried out on 2009 March 30 at 15:53 UT. 
	The IRCZ4 {\it AKARI} IRC observing template (AOTZ4) was used.
	The NG mode of IRCZ4 uses a near-IR grism (2.5--5 $\mu$m), in which 
	a target is located on the small $1\arcmin \times 1\arcmin$ aperture (Np)
	for point-source grism spectroscopy.
	The effective spectral resolution is $R \sim 100$ at 3.6 $\mu$m for a point source in Phase 3.
	The spectral resolution is expected to be lower than this value for extended sources such as comets.
	Raw data were processed through the IRC Spectroscopy Toolkit for Phase 3 data
	(version 20090211) with the new spectral responsivity (version 20091113)\footnote[9]{\footnotesize{See {\it AKARI (ASTRO-F) Observers Page} 
	(http://www.ir.isas.jaxa.jp/ASTRO-F/Observation/)}}.
	A one-dimensional spectrum of the C/Lulin was extracted with a pseudo aperture
	of 60\arcsec$ \times $4\farcs5 at $4\farcs5$ west 
	from the comet nucleus (see Figure \ref{Lulin_img}).
	Unfortunately, the east side coma to the nucleus is affected by the contamination of a background field star.
	We select this pseudo aperture position in order to avoid the opacity effect in the source
	(see below).
	The corresponding aperture position for the spectrum extraction is 
	depicted in Figure \ref{Lulin_img}.

	Imaging observations with AOTZ3 were performed on 2009 March 31 at 00:07 UT,
	which was 8 hr after the spectroscopic observations,
	yielding images of 1.9--2.8 (N2 band), 2.7--3.8 (N3), and 3.6--5.3 (N4) $\mu$m
	for at least two dithered positions.
	The reference wavelengths of N2, N3, and N4 bands are 2.4, 3.2, and 4.1 $\mu$m, respectively.
	The data were processed with the IRC Imaging Toolkit for Phase 3 data
	(version 20081015; see footnote 9).
	The resultant IRC image has a pixel size of $1\farcs46$
	and the FWHM of the image size is $\sim 4\farcs7$ in Phase 3 \citep{Onaka08},
	corresponding to 1440 km and 4630 km at the geocentric distance of 1.36 AU, respectively.
	
	The observation parameters are summarized in Table~\ref{journal}. 
	The temperature of the telescope system and the IRC during Phase 3 is above 40 K
	and is gradually increasing.
	Note that we 
	need further careful calibration and analysis with regard to the absolute flux calibration
	and weak spectral features,
	although any remarkable systematic changes in the sensitivity are not seen in Phase 3 
	at the moment.
	Please refer to 
	{\it AKARI} IRC Data User Manual for Post-Helium (Phase 3) Mission (see footnote 9)
	for more detail of the calibration in Phase 3.
	The performance in the LHe cryogen phase is described in IRC instrument papers
	\citep{Onaka07, Ita07, Ohyama07}.
	



\section{Results and Discussion}

	Figure \ref{Lulin_img} shows a RGB false color image of C/Lulin produced from the
	{\it AKARI}/IRC N2 (blue: 1.9--2.8 $\mu$m), N3 (green: 2.7--3.8 $\mu$m), and N4 (red: 3.6--5.3 $\mu$m) band data.
	The images are degraded by a Gaussian beam with the FWHM of $4\farcs5$,
	so that they match with each other in the point-spread function.
	The {\it AKARI} images of comet exhibit a crescent-like shape coma sunward and a dust tail 
	extending toward the anti-solar direction (i.e., eastward).
	The N4 (red) image (CO/CO$_2$ and dust grains) has a distribution different 
	from those of the N2 (blue) and N3 (green) images (H$_2$O and dust grains).
	Since both the N2 and N3 filters cover the 2.7--2.8 $\mu$m region of H$_2$O emission,
	the blue and green images resemble each other.
	
\begin{figure}
   \centering
   \includegraphics[scale=0.7]{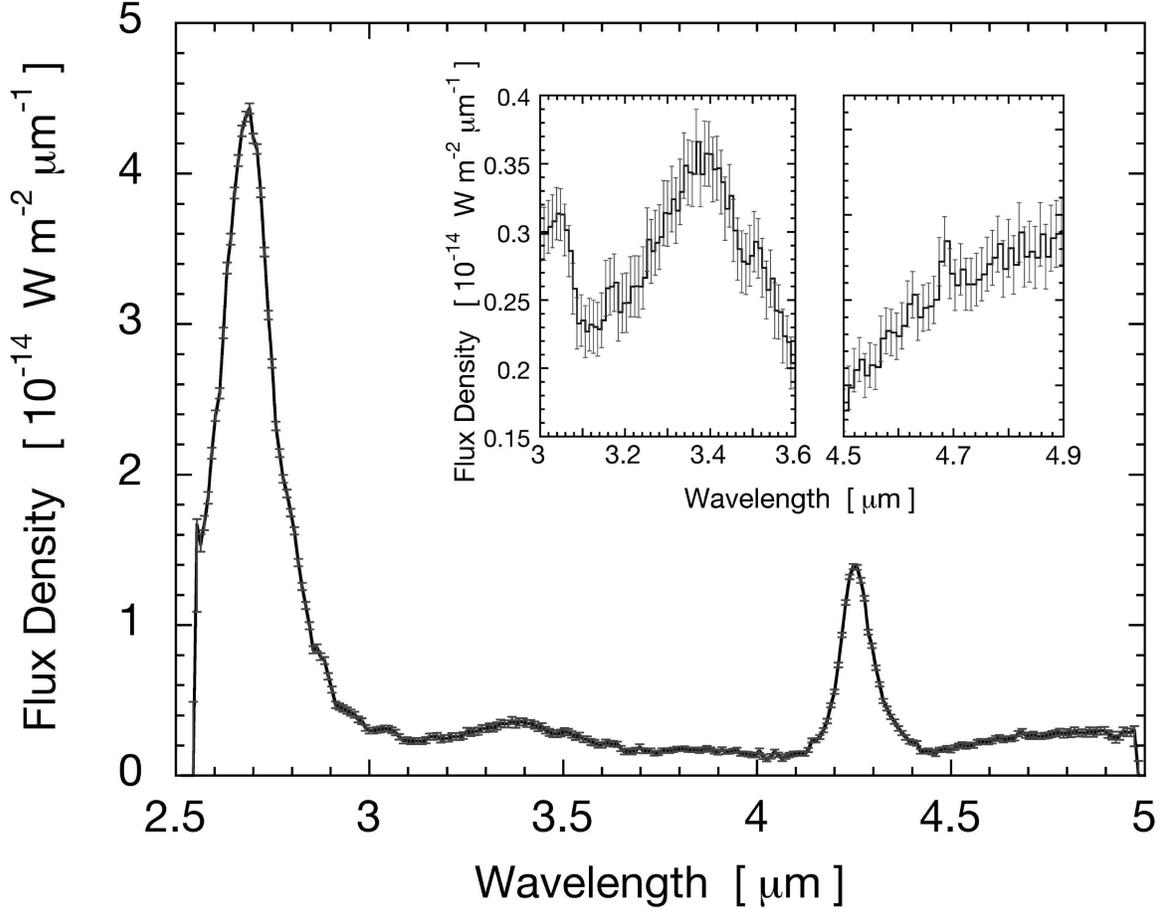}
      \caption{2.5--5 $\mu$m spectrum of comet C/2007 N3 (Lulin)
      obtained with the {\it AKARI}/IRC on 2009 March 30.
      The spectrum was extracted with the 60\arcsec$ \times $4\farcs5 aperture
      at $4\farcs5$ west from the comet nucleus (as shown in Figure \ref{Lulin_img}).
      Two distinct vibrational $\nu_3$ bands of H$_2$O at 2.66 $\mu$m and of CO$_2$
      at 4.26 $\mu$m are present in the spectrum.
      In addition to these two strong band features, the carbon monoxide $v$(1--0) 
      band at 4.67 $\mu$m and a broad 3.2--3.6 $\mu$m organic emission band can be seen.
      The inset figure shows the close-up spectrum in the 3--3.6 and 4.5--4.9 $\mu$m region.
        }
         \label{Lulin_spc_1}
   \end{figure}

	The {\it AKARI} spectrum of C/Lulin is characterized by typical vibrational
	bands of parent molecules in the coma (Figure \ref{Lulin_spc_1}).
	Two strong $\nu_3$ bands of H$_2$O at 2.66 $\mu$m and CO$_2$
	at 4.26 $\mu$m are present in the {\it AKARI} spectrum of C/Lulin.
	In addition, the carbon monoxide $v$(1--0) 
	band at 4.67 $\mu$m and a broad 3.2--3.6 $\mu$m emission band, which
	corresponds to a stretching mode of C--H in hydrocarbons, can be seen.
	We will concentrate on their interpretation based on their
	derived production rates (outgassing rates) in the following.

	To estimate the molecular abundance in the comet nucleus, 
	the flux densities are converted into the molecular production rates.  
	For the accurate measurement of the production rate, opacity effects should be 
	taken into account, because it is probable that these parent molecules are 
	optically thick near the nucleus \citep{Combes88, Crovisier06}.
	\citet{Feaga07} discussed this opacity effect near the nucleus of 9P/Tempel 
	for H$_2$O and CO$_2$, and derived the critical column density at which 
	opacity effects become non-negligible for two cases of the rotation temperature
	$T = 30$ and 100 K.
	Based on their estimate, 
	opacity must be considered within the projected distance of $\sim 20$ km from the nucleus
	for the production rates $Q\rm{(H_2O) \sim 5 \times 10^{27}}$ and 
	$Q\rm{(CO_2) \sim 3 \times 10^{26}}$ molecules\,s$^{-1}$. 
	The region more than 1000 km ($\sim 1\arcsec$) far from the nucleus is expected 
	to be almost free from the opacity effects in the case of our observation
	(C/Lulin was about 20 times brighter than 9P/Tempel).  
	We select the slit region as 60\arcsec$ \times $4\farcs5 ($40 \times 3$ pixels)
	at $4\farcs5$ west from the comet nucleus 
	to avoid the opacity effect (Figure \ref{Lulin_img}).
	In the following, we apply the method given in the Appendix of \citet{Hoban91} and \citet{BM04} 
	for the off-nucleus rectangular aperture 
	to calculate the production rates of parent molecules.
	We have considered an expansion velocity of 0.8 km\,s$^{-1}$ \citep{Combi88,Colangeli99}.
	The $g$-factors (fluorescence efficiency) and the molecular lifetimes are taken from 
	\citet{BMC89} and
	Jacques Crovisier molecular data base\footnote[10]{http://www.lesia.obspm.fr/perso/jacques-crovisier/basemole/}.

	In Figure \ref{Lulin_spc_1}, the scattered sunlight ($< 3$ $\mu$m) and the 
	thermal emission ($> 4$ $\mu$m)
	from the coma dust grains considerably contribute to the total flux of the continuum. 
	To derive the intensity of each molecular band, the continuum is subtracted and 
	the total band flux is integrated over the spectral range at 2.56--2.83, 4.1--4.4,
	and 4.64--4.72 $\mu$m for H$_2$O, CO$_2$, and CO, respectively.
	Table~\ref{bands} lists the intensities of
	these bands and the corresponding molecular production rates.
	Note that these are the values as of March 30.
	Periodic variations in the outgassing rate of C/Lulin are reported for many molecules \citep[e.g.,][]{Biver09}.
	The time and spatial variations are beyond the scope of this Letter and 
	will be studied in the forthcoming papers.
	
	The water band around 2.7 $\mu$m is a blend of at least two fundamental vibration
	($\nu_1$ and $\nu_3$) bands and hot bands ($\nu_2 + \nu_3 - \nu_2$ and $\nu_1 + \nu_3 - \nu_1$),
	but the $\nu_3$ band is expected to have the largest intensity and make 
	the most dominant contribution to this region.
	The weak OH $v$(1--0) resonance also contributes to the shoulder around 2.8 $\mu$m \citep{BMC89}.
	We integrated the water band flux in the 2.56--2.83 $\mu$m range.
	The contribution of weak OH $v$(1--0) band to the entire band flux is expected to be negligible. 
	We used the value $3.6\times10^{-4}$ s$^{-1}$ as the total H$_2$O $g$-factor in this region \citep{BMC89}.
	The observed band flux corresponds to a water production rate 
	$Q\rm{(H_2O)}=(7.6\pm0.1)\times10^{28}$ molecules s$^{-1}$
	(hereafter, the absolute calibration uncertainty is not included in the error).
	
	CO$_2$ cannot be directly observed in the near-IR from the ground because of
	strong telluric absorption.
	As described above, CO$_2$ has been directly detected only in four comets thus far.
	The $\nu_3$ band of CO$_2$ at 4.26 $\mu$m is clearly detected in our spectrum of C/Lulin.
	By assuming an expansion velocity of 0.8 km\,s$^{-1}$, we have derived, directly
	from our integrated band flux, the production rate 
	$Q\rm{(CO_2)}=(3.4\pm0.1)\times10^{27}$\,molecules\,s$^{-1}$.
	The relative production rate of $\rm{CO_2}$ compared to $\rm{H_2O}$ derived 
	from our observation is 4.5\%.  
	The ratio of the $\rm{CO_2}$ to $\rm{H_2O}$ production rates is greater than 
	20\% for comet Hale-Bopp at 2.9 AU \citep{Crovisier97b}
	and $\sim 8$\%--10\% for Hartley 2 at 1.0 AU \citep{Colangeli99, Crovisier99b}, respectively. 
	Comet Hale-Bopp was, however, observed at more than 2.9 AU far from the Sun.
	Its high ratio ($> 20$\%) is likely due to the high volatility of CO$_2$ 
	compared to H$_2$O.
	Using $Q\rm{(CO_2)}$/$Q\rm{(CO)}$ of $\sim 0.3$ obtained at 2.9 AU, 
	the ratio at 1 AU can be extrapolated as $\sim 6$\% \citep{BM04}.
	In the case of comet 9P/Tempel at 1.5 AU, it is reported that CO$_2$ to H$_2$O ratio 
	is $\sim 7$\% \citep{Feaga07}.
	Each Letter, however, adopted different values for the $g$-factor of water.
	In Colangeli et al. (1999) and Feaga et al. (2007), they considered a main contribution of $\nu_3$ band alone and 
	used the $g$-factor of $2.6 \times 10^{-4}$ and $2.85 \times 10^{-4}$ for H$_2$O, respectively.
	If the same value $3.6 \times 10^{-4}$ as this Letter is adopted for water, 
	we obtained $\rm{[CO_2/H_2O]}\sim 11$\% for 103P and $\sim 9$\% for 9P. 
	As for comet 73P, it is suggested that $Q\rm{(CO_2)}$/$Q\rm{(H_2O)} \sim 5$\%--10\% 
	by {\it Spitzer} imaging observations \citep{Reach09}.
	C/Lulin has a relative CO$_2$ abundance lower than Jupiter-family comets.

	The first clear detection of the near-IR $v$(1--0) band of CO around
	4.7 $\mu$m was made at observations of Comet Hyakutake \citep{Mumma96, DiSanti03}.
	The CO band in near-IR has been detected for eight Oort cloud comets and two Jupiter-family comets
	with high-dispersion spectrometers from the ground \citep{BM04, DiSanti08}.
	For Oort cloud comets observed with IR ground-based spectroscopy, 
	the total CO abundance ranged from 1\% to 24\% relative to H$_2$O
	\citep{Mumma03}.
	Carbon monoxide can be produced in the cometary coma from other precursors,
	and it can then exhibit both native (direct from the comet nucleus) and extended
	(or distributed) sources in comets.
	Both native and extended sources may contribute to this value.
	It is suggested that the native abundance CO/H$_2$O is found to be 0.4\%--17\% \citep{Mumma03, DiSanti08}.
	As for the two Jupiter-family comets (9P/Tempel and 73P/Schwassmann--Wachmann), 
	$\rm{CO/H_2O} \sim 4$\% and $< 3$\%, respectively \citep{DiSanti08}. 
	On the other hand, space and in situ missions could not detect this band clearly due to
	the low spectral resolution except for comet Hale-Bopp by {\it ISO} \citep{Crovisier99a}.
	In the {\it AKARI} spectrum of C/Lulin, a weak (2$\sigma$) band can be seen 
	around 4.68 $\mu$m (Figure \ref{Lulin_spc_1}).
	It is most likely that this corresponds to the 4.67 $\mu$m CO $v$(1--0) band.
	The comet Halley spectrum by {\it Vega} ($R\sim85$) might resolve the P and R branches of CO, 
	although the CO band strength is near the detection limit \citep{Combes88}.
	The resolved P and R branches are not expected in the IRC spectrum
	because of the spectral resolution $\sim 100$ \citep[see Figure 7 in][]{Crovisier87}.  
	Our derived upper limit of the CO production rate is $Q\rm{(CO)} < 1.3 \times 10^{27}$ molecules\,s$^{-1}$ (3$\sigma$),
	when we integrate the flux in the 4.64--4.72 $\mu$m region.
	Improvements in the calibration will lower the detection limit for faint features.
	It is secure that the corresponding production rate of $\rm{CO}$ compared 
	to $\rm{H_2O}$ is less than 2\%.
	
	The 3.52 $\mu$m methanol (CH$_3$OH) feature can also be seen upon a broad 3.2--3.6 $\mu$m emission 
	band from C--H bearing molecules in the spectrum (Figure \ref{Lulin_spc_1}).
	These 3.2--3.6 $\mu$m features may be attributed to a blend of many rovibrational lines 
	belonging to methanol, organics, and hydrocarbons such as methane (CH$_4$) and ethane (C$_2$H$_6$) \citep{BM95}.
	Methane, ethane, and methanol have been observed in many comets in the near-IR
	with the ground-based high-resolution spectroscopy \citep[][and references therein]{DiSanti08}.  
	Further detailed identification of each feature in the {\it AKARI} spectrum
	is difficult at the present calibration stage.
	
	The relative abundances $\rm{CO_2/H_2O}$ and $\rm{CO/H_2O}$ derived from our observations are
	$\sim4$\%--5\% and $< 2$\%, respectively.
	The value $Q\rm{(CO_2)}$/$Q\rm{(H_2O)} \sim 4$\%--5\% is similar to 1P/Halley \citep[3\%--4\%;][]{Combes88}
	and the estimated value ($\sim 6$\%) of Hale-Bopp at 1 AU \citep{BM04}.
	The value $Q\rm{(CO)}$/$Q\rm{(H_2O)} < 2$\% of C/Lulin belongs to the group that has relatively low
	$Q\rm{(CO)}$/$Q\rm{(H_2O)}$ values among the comets observed ever.
	It is similar to the value 0.9\% of Oort comet C/1999 S4 \citep[LINEAR;][]{Mumma01}.
	C/Lulin was observed at 1.7 AU of the heliocentric distance after the perihelion passage.
	Since CO is one of the highly volatile species in comet nuclei,
	CO might be exhausted from the surface of the comet nucleus
	and highly depleted at the observation epoch, although such phenomena have not yet been reported.
	It is more probable that the depletion occurred if the comet nucleus condensed 
	at moderately high nebular temperatures.
	In the present paradigm, comet formation in the early solar nebula extended over a wide
	range of the heliocentric distance both for Oort cloud and Jupiter-family comets
	\citep{Dones04, Duncan04}.
	This suggests that comets could display diversity in their chemical composition depending on
	the local temperature and nebular composition where they are formed \citep{BM04}. 
	It is believed that Oort cloud comets were formed 
	in the giant planet region of the early solar nebula, although recent dynamical models 
	suggest that comets now in the Oort cloud were contributed in roughly equal numbers both from
	giant planet and the Kuiper Belt region \citep{Dones04}.
	It is reasonable to infer that C/Lulin has a small fraction of CO and CO$_2$ in its nucleus
	because it may originate from the region closer to the Sun among the planetesimal formation cites.
	
	Establishing a chemical taxonomy for comets gives us important insights into the planetesimal 
	formation process in the early solar nebula, and it is important to study the chemical 
	diversity of the parent molecules in many more comet samples. Near-IR observations with {\it AKARI} 
	will provide precious data for these parent molecules.


\section{Conclusions}

	We reported the near-IR spectrum and image of comet C/2007 N3 
	(Lulin) observed with {\it AKARI}/IRC on 2009 March 30 and 31, respectively.
	Both the fundamental vibrational bands of water (2.66 $\mu$m) and carbon dioxide (4.26 $\mu$m)
	are clearly detected in the {\it AKARI} spectrum. In addition, 3.2--3.6 $\mu$m
	C--H bearing molecule feature and $v$(1--0) band of carbon monoxide (4.67 $\mu$m) are also
	detected. The relative abundances $\rm{CO_2/H_2O}$ and $\rm{CO/H_2O}$ 
	derived from our observations are $\sim 4$\%--5\% and $< 2$\%, respectively.
	Comet Lulin belongs to the group that has relatively low abundance
	of CO and CO$_2$ among the comets observed ever.



\acknowledgments

	This work is based on observations with {\it AKARI}, a JAXA project with the participation of ESA.
	We thank all members of the {\it AKARI} project for their continuous help and support.  
	T.O. and S.H. were supported by the Space Plasma Laboratory, ISAS, JAXA.
	This work is supported in part by Grant-in-Aid for Young Scientists (B)
	No. 21740153 (T.O.) from the Ministry of Education, Culture, Sports, Science and 
	Technology of Japan.



{\it Facilities:} \facility{AKARI(IRC)}.

\clearpage

\begin{table}
\caption{Observational Summary}
\label{journal}
\renewcommand{\footnoterule}{}  
\begin{tabular}{ccccc}
\hline
UT Date             &  AOT  & Heliocentric & Geocentric & Phase  \\
                    &       & Distance     & Distance   & Angle  \\
                    &       & (AU)         & (AU)       & (deg)  \\
\hline
2009 Mar 30 15:53 & IRCZ4(spectroscopy) & 1.70 & 1.36 & 36.0 \\
2009 Mar 31 00:07 & IRCZ3(imaging)      & 1.70 & 1.37 & 35.9 \\
\hline
\end{tabular}
\end{table}
\clearpage
\begin{table}
\caption{Molecular Bands Observed in C/2007 N3 (Lulin) with {\it AKARI}/IRC}
\label{bands}
\begin{tabular}{ccccccc}
\hline
Molecule & Band & $\lambda$ & $g$\tablenotemark{a} & $\lambda\lambda$\tablenotemark{b} & Flux\tablenotemark{c} & $Q$\tablenotemark{d}\\
         &      & ($\mu$m)  & (s$^{-1}$)         & ($\mu$m)                          & (W m$^{-2}$)          & (s$^{-1}$) \\
\hline
\multirow{4}*{H$_2$O} & $\nu_3$ & 2.66   & $2.82\times10^{-4}$ & \multirow{4}*{2.56--2.83} & \multirow{4}*{$(4.42\pm0.03)\times10^{-15}$} & \multirow{4}*{$(7.6\pm0.1)\times10^{28}$} \\
       & $\nu_1$ & 2.73   & $0.25 \times 10^{-4}$ &             &  & \\
       & $\nu_2+\nu_3-\nu_2$ & 2.66 & $0.28 \times 10^{-4}$ &   &  & \\
       & $\nu_1+\nu_3-\nu_1$ & 2.73 & $0.22 \times 10^{-4}$ &   &  & \\
CO$_2$ & $\nu_3$ & 4.26 & $2.9 \times 10^{-3}$ & 4.1--4.4   & $(1.23\pm0.02) \times 10^{-15}$ & $(3.4\pm0.1) \times 10^{27}$\\
CO     & $v$(1--0) & 4.67 & $2.6 \times 10^{-4}$ & 4.64--4.72 & $< 4.0\times 10^{-17}$ &  $< 1.3\times 10^{27}$\\
\hline
\end{tabular}
\tablenotetext{a}{Emission rate assuming resonant fluorescence exited by the Sun at 1 AU}
\tablenotetext{b}{Wavelength integration range where the fluxes are computed}
\tablenotetext{c}{Absolute calibration uncertainty is not included in the error}
\tablenotetext{d}{Production rate assuming a molecule distribution with an expansion velocity of 0.8 km\,s$^{-1}$}
\end{table}

\clearpage



\clearpage

\end{document}